\documentclass[a4paper,fleqn]{cas-dc}

\usepackage[numbers]{natbib}
\usepackage{algorithm}
\usepackage{algpseudocode}
\usepackage{algorithm}
\usepackage{graphicx}
\usepackage{subcaption}
\usepackage{longtable}
\usepackage{booktabs}
\usepackage{float}
\usepackage{stfloats}
\usepackage{listings}
\usepackage{xcolor}

\begin{document}

\let\WriteBookmarks\relax
\def\floatpagepagefraction{1}
\def\textpagefraction{.001}

% -------- TITLE --------
\shorttitle{Flow-Graph-Driven Multi-Agent Framework}
\shortauthors{Srita et al.}

\title [mode = title]{FGDM: Reasoning Aware Multi-Agentic Framework for Software Bug Detection using Chain of Thought and Tree of Thought Prompting}

% -------- AUTHORS --------
\author[1]{Srita Padmanabhuni}
\ead{srita_padmanabhuni@srmap.edu.in}

\author[1]{Bhargavi Karuturi}
\ead{bhargavi_karuturi@srmap.edu.in}

\author[1]{Jerusha Karen Indupalli}
\ead{jerushakaren_indupalli@srmap.edu.in }

\author[1]{Santhan Reddy Chilla}
\ead{santhanreddy_chilla@srmap.edu.in  }

\author[1,2]{Vivek Yelleti}
\ead{vivek.yelleti@gmail.com}

\affiliation[1]{organization={SRM University, Andhra Pradesh},
    city={Amaravati},
    country={India}}

\affiliation[2]{organization={Corresponding Author}}

% -------- ABSTRACT --------
\begin{abstract}
Deep Learning methods are becoming prominent in automated software bug detection; however, they lack the global understanding of the given code. Consequently, their performance tends to degrade, especially when they are applied to large interconnected code bases or complex modular programs. Recently, Large Language Models (LLMs) have proven to be effective at capturing dependencies among multiple interconnected modules in the codebase. This motivated us to propose the Flow-Graph-Driven Multi-Agent Framework (FGDM), which is composed of four agents that operate in a sequential manner. The framework converts the received code to a flow graph, identifies the erroneous segments, and further generates the repaired code. All the employed agents utilize Chain-of-Thought (COT) and Tree-of-Thoughts (TOT) prompts. Additionally, we also integrated with the FAISS vector database to retrieve similar previous bugs and their repairs. We demonstrated the efficacy of the proposed framework over 100 programs from several projects, including Ansible, Black, FastAPI, Keras, Luigi, Matplotlib, Pandas, Scrapy, SpaCy, and Tornado in both C and Python programs. Our experiments demonstrate that the FGDM outperforms the extant approaches and yielded reductions with a mean of 24.33 and 8.37 in Levenshtein distance and similarities of 0.951 and 0.974 in cosine similarity for Python and C, respectively. 
\end{abstract}

% -------- KEYWORDS --------

\maketitle

% -------- INTRODUCTION --------
\section{Introduction}
Often software systems are prone to failure because of the presence of software bugs \cite{10.1145/3750040,shiri2024systematic} in the code, which were kept, knowingly or unknowingly. This leads not only to incorrect outputs but also to significant operational losses. These defects in the software systems primarily occur due to oversight during the design process, a lack of proper understanding of the underlying system, misinterpreted coding errors, or issues that are overlooked during the integration of sub-modules. Hence, detecting these bugs at the early stage would save millions of dollars for the companies and also increase trust among the stakeholders. However, the major problem arises when the systems start to grow, leading to scalability issues. Hence, manual debugging has become almost impossible and is a highly time-consuming task, thereby necessitating the design of intelligent and automated software bug-detecting systems. 

In the early days, to automate the bug detection, researchers had designed the rule-based systems. The rules are designed in such a way that they cover the most common ways of the occurrence of bugs. These methods include the static analysis techniques and manual inspection too. Even though these methods turned out to be effective in specific scenarios, their major downside is their limited applicability over large code bases. This laid the path to design the machine learning (ML) \cite{shiri2024systematic} or deep learning (DL)-based architectures \cite{yang2022survey} where the classifiers such as decision tree, random forest, convolutional neural networks (CNNs), and recurrent neural networks (RNNs) are employed for the software bug detection. Various feature extraction processes extract features from the given code in the initial stage. Typically, all these ML/DL approaches consider the provided source code as a linear sequence or text. These methods limited their ability to capture structural relationships and dependencies \cite{ahmed2023deep} across different parts of a program. This results in decreasing the effectiveness of these approaches over the large, interconnected, or non-modular software systems.

The fundamental limitation of many existing approaches lies in their lack of context-aware bug detection. As explained earlier, traditional approaches where ML/DL techniques are employed analyze the code line by line or as isolated code blocks. This approach limits the models, restricting their understanding only to capture the local context, ignoring the global context. In the real world, software bugs primarily arise from misalignment with the global needs of the software systems. In other words, the software developers design the code in an isolated environment, but when they get integrated, the major objective of theirs will get disengaged with the needs of the software system. Hence, while detecting the software bugs, considering the contextual reasoning also into the consideration is considered important. Context-aware bug detection aims to capture these interdependencies by considering control flow, data flow, and cross-module relationships within the entire program. This global perspective enables the identification of deeper logical errors, such as inconsistent state propagation, incorrect function interactions, and broken dependencies, which are often missed by localized analysis. Therefore, incorporating structural and contextual information is critical for improving the accuracy and robustness of automated debugging systems.

Recently, introduction of Large language models (LLMs) have opened up new opportunities and laid the path towards the intelligent software engineering \cite{zhu2025software}. LLMs can effectively understand the programming languages and can capture contextual dependencies between multiple modules. However, LLMs suffers majorly from the following limitations, e.g. hallucination, which means that the model generates outputs that sound plausible but are semantically incorrect/inconsistent with the real program logic. Ultimately, due to these hallucinations based predictions by LLMs could lead to incorrect bug localization or incorrect code repairs, creating new bugs instead of fixing existing ones. Also, LLMs can have unstable reasoning, lack traceability in decision-making and be sensitive to prompt design.

To deal with these problems, structured reasoning strategies such as Chain-of-Thought (CoT) \cite{wei2022chain} and Tree-of-Thoughts (ToT) \cite{yao2023tree} prompting have been proposed. CoT encourages the model to generate intermediate reasoning steps, which leads to more transparency and less spuriously supported conclusions. ToT extends this idea by considering multiple reasoning paths and choosing the best consistent solution, which is especially useful in difficult debugging tasks with several potential fixes. While these strategies do not completely remove hallucination, they help mitigate it, by enforcing more grounded and systematic reasoning. Hence, it is important to embed these reasoning methods in a well-structured framework to guarantee dependable performance.

A further important limitation of the present literature is the absence of language independent solutions. Most of the existing approaches are designed and optimized for specific programming languages, e.g., C or Python, and are therefore not usable in heterogeneous software ecosystems. This limitation is often caused by language-specific representations, handcrafted rules, or prompt designs that fail to generalize across program paradigms. In fact, modern software systems often mix different languages, so it's important to develop flexible, language-agnostic solutions. A language-independent framework not only improves generalizability, but also frees up the burden of re-designing models and prompts for every programming language, making them more scalable and usable for real-world applications.

In this paper, we propose a Flow-Graph-Driven Multi-Agent Framework (FGDM) for automated bug detection and fixing to address these challenges in this paper. This approach represents any given  program as a flow graph so as to preserve the structural and control-flow information to enable effective context aware analysis. The framework utilizes four expert agents in a pipeline manner to conduct graph construction, bug localization, code repair and source code reconstruction. This multi-agent design diminishes dependence on a single reasoning process, thus improving robustness and reliability. Both CoT and ToT prompting strategies are used for each agent to boost reasoning abilities and alleviate hallucination effects. Moreover, we add an FAISS-based vector retrieval component that retrieves similar historical bugs and their fixes for retrieval-augmented reasoning. The use of flow graph representations further helps in language-independent processing as the program structure is abstracted beyond syntax, thus enabling the framework to be adaptable across different programming languages.

The main contributions of this paper are summarized as follows:
\begin{itemize}
    \item A flow-graph-based approach for automatic bug localization and repair in complex Python and C programs.
    \item A multi-agent framework that systematically performs graph construction, fault localization, repair, and code reconstruction.
    \item Integration of Chain-of-Thought and Tree-of-Thought reasoning strategies to enhance debugging accuracy.
    \item Incorporation of a FAISS-based retrieval mechanism for leveraging historical bug-fix patterns.
    \item Comprehensive evaluation on real-world software projects demonstrating improved performance over baseline methods.
\end{itemize}

The rest of the paper is organized as follows: Section 2 present the
preliminaries that are required for this study, and Section 3 overview of the extant related works. Section 4 presents the proposed framework in a detailed manner. Section 5 presents experimental results and performance evaluation. Section 6 concludes the paper.

% -------- PRELIMINARIES --------
\section{Preliminaries}

\subsection{Prompt Engineering}
Large Language Models (LLMs) are trained with huge volumes of datasets corresponding to diverse domains such as engineering, biology, finance, health care, etc. Hence, they can be utilized to handle various kinds of datasets. Prompt engineering is an art and science with which one can enable getting the desired output from the LLMs. This helps in designing and optimizing prompts and helps to guide LLMs. Since the quality of the prompts plays a critical role in the quality of the responses, careful crafting of the prompting is of the utmost importance.

\subsection{Chain-of-Thought}
Chain-of-Thought (COT) is a kind of prompting technique that directs the LLMs to generate better-quality output. It involves breaking down complex tasks into simple steps and mimicking human thinking, thereby facilitating multi-step thinking and incorporating step-wise thinking. By designing the prompts with CoT reasoning, LLMs solves the given problem by providing the step-by-step reasoning with more coherent logical steps. 

\subsection{Tree-of-Thought}
Tree-of-Thought (TOT) is also an advanced prompting technique like CoT, thereby improving the reasoning-based prompts. The main difference between CoT and ToT is it the way the logical steps are being stored. In ToT-based prompts, the logical steps are stored in the form of a tree. While solving, LLMs will be evaluating these intermediate thoughts through a deliberate reasoning process. Further, it utilizes search algorithms such as breadth-first search or depth-first search to generate and evaluate thoughts, thereby enabling the systematic exploration of thoughts with lookahead and backtracking.

\subsection{Flow Graph}
The flow graph model uses nodes and edges to depict a program, maintaining its structure and control flow. It comprises nodes, and they are interconnected to each other. Here, the nodes represent the independent code blocks, and the dependency between distinct code blocks is well captured by the relationships. This relationships can be either data flow or control flow. These flow graphs assist in identifying bugs more accurately than plain text models.

% --------LITERATURE REVIEW --------
\section{Literature Review}
Srita et al. \cite{11418178} proposed a meta-agentic framework, which is driven by various agents analyzing within the module, and across the module, where bug detection is performed. Further, they employed RAG to reports bug based on the historical knowledge attained thus far. In the second stage, they employed a meta agent that collates all the bugs detected by these modules and cross-verifies overall. This is applied to modular-based programs and is applied to Python programs. Ferrag et al. \cite{10910240} proposed SecureFalcon, with 121 million parameters and derived from the Falcon-40B model. It is explicitly tailored for classifying software vulnerabilities. They reported that the proposed model outperformed several other LLMs, machine learning, and deep learning models. Li et al. \cite{10.1145/3649828} proposed LLIFT, which turned out to be in the integration of static analysis and user before initialization. This approach utilizes the conventional constraints present in Linux and analyzes the bugs using post-constraint guidance. Zhang et al. \cite{10232867} conducted an extensive analysis and reported that the pretrained models outperform extant approaches. Jiang et al. \cite{9782533} proposed BugBuilder, with which they automated the process of constructing bug repositories by utilizing version control systems. Guan et al. \cite{guan2025crossprobe} proposed an LLM-driven approach that utilizes the bugs detected in other programming languages to detect bugs in other languages. In this way, they are incorporating the cross-domain knowledge. Sijwali et al. \cite{sijwali2025fixing} proposed an LLM-driven framework to automatically detect and repair the errors in Java programs. They employed the GPT-4o-mini model, which is utilized for contextual signals—including code diffs, developer comments, and bug reports. 

Nowadays, smart contracts are becoming the standard operating procedure for the financial transactions. Hence, a small bug in them could potentially lead to huge economic loss. Zhang and Zhang \cite{zhang2024detecting} proposed a hybrid system, where they combined both LLM and rule-based reasoning for the identification of error vulnerabilities in smart contracts. Hai et al. \cite{hai2022cloud} employed multi-layer perceptron (MLP) for the identification of bugs in the cloud-based bug tracking software. They employed various deep learning architectures, and MLP turned out to be the best-performing model.  Pradel et al. \cite{pradel2018deepbugs} proposed DeepBugs, which automatically learns the patterns and identifies the bugs based on the name-based detection process. Further, it analyzes the semantic representations along with the syntactic representations. Kukkar et al. \cite{kukkar2020duplicate} employed convolutional neural network (CNN) for the identification of duplicate bug report detection. Initially, CNN captures all the relevant features, and then the duplicate reports are captured. The authors reported that CNN yielded an accuracy of 85-99\%. Pham et al. \cite{pham2019cradle} proposed CRADLE, which operates in two staged approach: (i) initially, the inconsistency in the cross-implementation is analyzed; and (ii) anomaly propagation tracking is employed to localize the faulty functions. They evaluated the performance of the proposed approach over the DL libraries such as TensorFlow, CNTK, and Theano. Xi et al. \cite{xi2019bug} proposed SeqTriage, a deep learning-based approach that employs long short-term memory (LSTM) and processes the code line by line. They integrated features collected from textual content, metadata, and tossing sequences. 

Zheng et al. \cite{zeng2024classifying} proposed a transformer-based method for predicting bug severity. They fine-tuned the model with domain-specific pretraining strategies. On similar lines, Fukuda et al. \cite{fukuda2022fault} employed Transformer models to generate comprehensive human-readable fault reports. They showed the effectiveness of the model over heterogeneous data sources. Wei et al. \cite{wei2023improving} also proposed Transformer-based techniques to enhance the extraction of security domain knowledge, thereby improving the relevance and quality of extracted keywords. Zhang et al. \cite{10.1145/3699598} employed LLMs for generating the unit assertions which assisted to detect the bugs in real-time environments.

Siva et al. \cite{siva2024automatic} proposed an evolutionary algorithm-driven deep learning-based methodology for software bug detection. It operates in the following manner: (i) initially, the features are extracted by using the adaptive golden eagle optimizer; (ii) thereafter, these features are altered by invoking the opposition-based learning. These processed features are passed to Long Short-Term Memory (LSTM) and Recurrent Neural Network (RNN) for bug detection. Mostafa et al. \cite{MOSTAFA2025112205} proposed the utilization of a weighted genetic algorithm, which is applied over the extracted features. Later, these features are transferred to any predictive models (either machine learning / deep learning) based models for the identification of software bug detection. In a similar way, Cynthia et al. \cite{10.1145/3511430.3511444} employed the weighted genetic algorithm for transforming the features and then employed KNN or Random Forest (RF) for the identification of software bugs. Juneja et al. \cite{Juneja2025} proposed Automatic Duplicate and Learning-based Model (ADLM model). This model operates in a three-stage manner as follows: (i) in the first stage, all the preprocessing methods are employed to be invoked; (ii) in the second stage, features are extracted at multiple levels, including temporal, categorical, contextual, and textual features. (iii) All the obtained features from the second stage are then combined and processed by the LSTM network to identify the software bugs in the given programs. Garg et al. \cite{Garg2024} proposed a modified deep neural network (MDNN) which works as follows: (i) after preprocessing the extracted features, principal component analysis (PCA) is employed to reduce the dimensions. Thereafter, Fuzzy C-means Clustering Method (FCM) is invoked to extracted the correlation features. Then, the classifier is employed to predict the software bugs and the optimal weights are selected by the reptile search algorithm (RSA). Al-Fraihat et al. \cite{10477341} employed ensemble learning for the hyper parameter optimization and used it for software bug detection. Garg et al. \cite{10.1145/3540250.3549096} proposed deeplearning-perf, where they employed a transformer-based architecture for software bug detection. Tamewar et al. \cite{TAMESWAR2022100105} proposed hybrid Deep Neural Network model thereby enhancing the quality of the predicting the software bugs. They employed nature-inspired algorithm for getting the optimal hyperparameters. Bharath et al. \cite{10449266} employed logistic regression and random forest to anlayze the reports and then identify the software bugs. Dhruv \cite{11233621} employed various machine learning and deep learning models such as Random Forest (RF), Gradient Boosting (GB), and deep learning models like Convolutional Neural Networks (CNN) for the early identificaiton of software bugs. 

Recently, quantum has been gaining popularity due to the robustness of its security guarantees. Zhand and Miranskyy \cite{10821136} proposed a novel method to detect the falkyness in the reports generated from the quantum software. Nadim et al. \cite{Nadim2025} proposed the utilization of quantum support vector machine for the software bug detection.
\begin{figure*}
    \centering
    \includegraphics[width=1.0\linewidth]{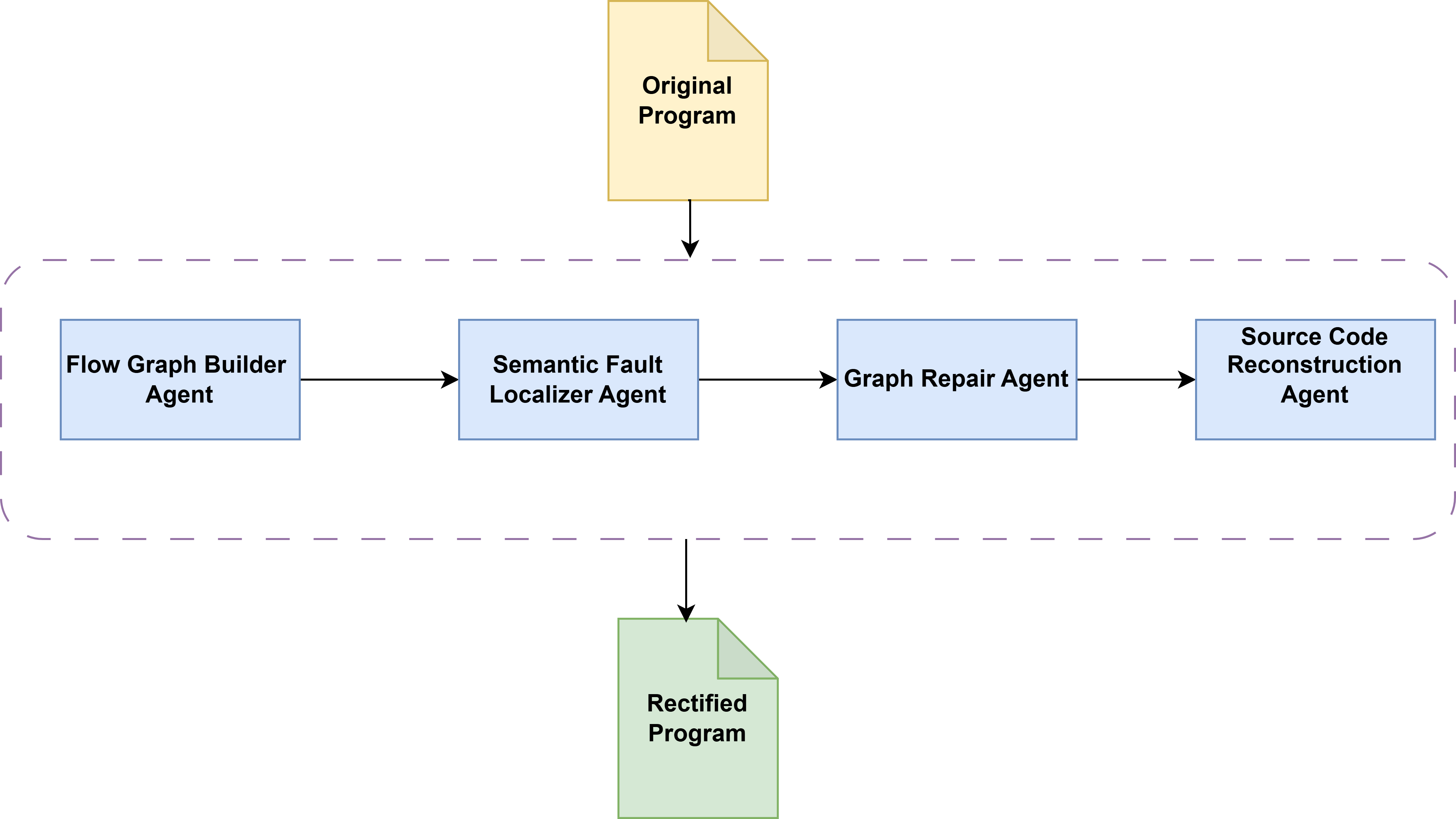}
    \caption{Schematic diagram of the proposed approach}
    \label{fig:placeholder}
\end{figure*}

% -------- METHODOLOGY --------
\section{Proposed Methodology}
In this section, we present the proposed Flow-Graph-Driven Multi-Agent Framework (FGDM) for software bug detection and repair. 

\subsection{Overview of the proposed Methodology}
The framework consists of four specialized agents operating in a sequential pipeline, enhanced by either Chain-of-Thought (COT) or Tree-of-Thought (TOT) prompting strategies. The overall workflow is summarized in Algorithm 1. Given buggy code, the proposed method invokes the following steps in a sequential manner: (i) all the required agents and the vector database are initialized in the first step; (ii) in the second step, the agentic pipeline is invoked, where it triggers the following four agents, namely: (a) flow graph builder agent: converts the given source code to the flow graph, (b) semantic fault localizer agent: identifies the faulty nodes, (c) graph repair agent: repairs and restructures the code only on the fault nodes, (d) source code reconstruction agent: generates the repaired code and gives recommendations to the users; (iii) the rectified code is also represented in the flow graph, so now the semantic validation is done in this step based on the employed graph correction criteria; (iv) later, the Levenshtein distance and cosine similarity between the given source code and the rectified code are computed and reported thereof. 
 
Now, we will discuss each and every step in a detailed manner as follows: 

\textbf{Step 1: Initialization of Agents and Vector Database} It starts with the initiation of the four dedicated agents: Flow Graph Builder Agent (Agent 1), Semantic Fault Localizer Agent (Agent 2), Full Graph Repair Agent (Agent 3), and Source Code Reconstruction Agent (Agent 4). The agents are set up to play specific roles in the debugging process, ensuring that the entire process proceeds
in an organized manner. Furthermore, the vector database using
FAISS algorithm is used in the framework to maintain previously analyzed flow graph structures along with the relevant bugs and fixes. This database acts as a knowledge base for all the agents. Using the stored similar graph structures and fix history, this process provides the means for retrieval-augmented reasoning throughout the pipeline, thus making the fault detection and fixing more effective through pattern matching.

\textbf{Step 2: Invoking Agentic Pipeline} The following four agents Flow Graph Builder Agent (Agent 1), Semantic Fault Localizer Agent (Agent 2), Full Graph Repair Agent (Agent 3), and Source Code Reconstruction Agent (Agent 4) are invoked in a sequential 

\textbf{Flow Graph Builder Agent} Given the source code, the following agents are invoked in a sequential manner. In the first stage, Flow Graph Builder Agent is invoked to convert the given source code to the flow graph. This agent utilizes the abstract syntax tree (AST) analysis, and is driven either with a Chain of Thought-based prompt or a Tree of Thought-based prompt. Here, the nodes represent the independent code blocks, and the dependency between distinct code blocks is well captured by the relationships such as (i) containment, (ii) data flow, (iii) control flow, and (iv) call edges.

\textbf{Semantic Fault Localizer Agent} It is important to note that the flow graph only consists of the nodes and relationships, but it does not represent whether the particular code block has a bug or not. Identifying this is done through the invocation of the Semantic Fault Localizer Agent. It analyzes the graph by using reasoning and other semantic evaluations to detect faulty nodes. Further, this agent also identifies the inconsistencies, such as broken dependencies and flow mismatches. Since, the code block consists of set of code lines, if atleast one of them is turned out to be having faulty then the entire code block is tagged to be as faulty node. 

\textbf{Graph Repair Agent} The role of this agent is to perform and suggest the structural corrections to be done over the faulty node and make it non-faulty. While doing so, the prompt is designed in such a way that it narrows down the faulty lines in the faulty node and the recommendations to be suggested to be as minimal as possible. In the end, it generates a rectified flow graph while preserving valid structures. This agent makes sure that only the affected lines are changed and does the following validations:
\begin{itemize}
\item \textbf{Structure Preservation:} All original vertices are retained, while only a minimal number of edges are modified.
\item \textbf{Defect Coverage:} All defective vertices identified by Agent~2 are addressed in the rectification process.
\item \textbf{Minimal Edge Manipulation:} The number of added or modified edges is less than or equal to the number of defective vertices, i.e., $|\text{modified edges}| \leq |\text{defective vertices}|$.
\end{itemize}

\textbf{Source Code Reconstruction Agent} This agent major task is to convert the repaired flow graph to the rectified code. Further, once the recitfied code is generated, it rechecks once again for the faulty lines and recommends the changes. 

All the above four agents either invoke CoT or ToT-based prompts based on the pipeline that is chosen to execute. All these prompts utilized for the CoT and ToT pipelines are presented in the Supplemental File.

\textbf{Step 3: Reporting results}
Finally, the comparison will be done between the given source code and the rectified code and the results of Levenshtein Distance and Cosine
Similarity measures are reported thereof.

\begin{algorithm}[htbp]
\caption{Graph-Based Multi-Agent Code Analysis}
\begin{algorithmic}[1]
\label{alg:framework}

\Require Source files $F = \{f_1, f_2, \dots, f_n\}$, LLM $L$, prompts $P_1, P_2, P_3, P_4$, output folders $G_1, G_2, G_3, G_4$
\Ensure Fixed files $F_{\text{fixed}}$, metrics $M$, graph sets $G_1, G_3$

\State \textbf{// Agent 1: Source Code $\rightarrow$ Flow Graphs}
\State $G_1 \gets \emptyset$
\For{each $f_i \in F$}
    \State $code \gets \text{read}(f_i)$
    \State $result \gets \text{analyze\_code}(code, L, P_1)$
    \If{$|result.\text{nodes}| < 3$}
        \State $result \gets L(code + P_1^{\text{CoT}})$
    \EndIf
    \State $g_i \gets \text{build\_graph}(result)$
    \State $\text{write\_dot}(g_i, G_1/f_i)$
    \State $G_1 \gets G_1 \cup \{g_i\}$
\EndFor

\Statex

\State \textbf{// Agent 2: Fault Detection}
\State $G_2 \gets \emptyset$ \Comment{Storage for diagnoses}
\For{each $g_i \in G_1$}
    \State $stats \gets \text{compute\_graph\_stats}(g_i)$
    \State $prompt \gets P_2^{\text{ToT}}(stats)$
    \State $diagnosis_i \gets L(prompt)$
    \State $\text{write\_json}(diagnosis_i, G_2/f_i)$
    \State $G_2 \gets G_2 \cup \{diagnosis_i\}$
\EndFor

\Statex

\State \textbf{// Agent 3: Graph Repair}
\State $G_3 \gets \emptyset$
\For{each $i \in \{1, \dots, n\}$}
    \State $faulty\_details \gets \text{format\_faults}(diagnosis_i)$
    \State $full\_dot \gets \text{clean\_dot}(g_i)$
    \State $prompt \gets P_3(faulty\_details, full\_dot)$
    \State $repair_i \gets L(prompt)$
    \State $\text{write\_json}(repair_i, G_3/f_i)$
    \State $\text{write\_dot}(repair_i, G_3/f_i)$
    \State $G_3 \gets G_3 \cup \{repair_i\}$
\EndFor

\Statex

\State \textbf{// Agent 4: Code Reconstruction}
\State $F_{\text{fixed}} \gets \emptyset$
\For{each $i \in \{1, \dots, n\}$}
    \State $prompt \gets P_4(f_i, repair_i)$
    \State $fixed\_code_i \gets L(prompt)$
    \State $\text{write}(fixed\_code_i, F_{\text{fixed}}/f_i)$
    \State $F_{\text{fixed}} \gets F_{\text{fixed}} \cup \{fixed\_code_i\}$
\EndFor

\Statex

\State \textbf{// Similarity Metrics}
\State $M \gets \emptyset$
\For{each $i \in \{1, \dots, n\}$}
    \State $d_{\text{Lev}} \gets \text{levenshtein}(f_i, fixed\_code_i)$
    \State $d_{\text{line}} \gets \text{line\_dist}(f_i, fixed\_code_i)$
    \State $sim_{\text{cos}} \gets \text{cosine\_sim}(f_i, fixed\_code_i)$
    \State $M \gets M \cup \{(d_{\text{Lev}}, d_{\text{line}}, sim_{\text{cos}})_i\}$
\EndFor

\State \text{write\_json}($M$, \texttt{"results/output.json"})
\State \Return $F_{\text{fixed}}, M, G_1, G_3$

\end{algorithmic}
\end{algorithm}

% -------- RESULTS & DISCUSSIONS--------
\section{Results and Discussion}

In this study, the benchmark programs are selected from the BugsInPy \cite{Widyasari2020BugsInPy} repository, which contains real-world Python programs with annotated bugs. A total of 100 programs were used, including Ansible, Black, FastAPI, Keras, Luigi, Matplotlib, Pandas, Scrapy, SpaCy, and Tornado. We converted these programs into corresponding C version and utilized them in our research work.

These programs exhibit inherently flat and non-modular programs, where each file contains complex control and data dependencies. Such characteristics make them well-suited for graph-based representation and analysis without requiring artificial modularization. To evaluate the effectiveness of the proposed Flow-Graph-Driven Multi-Agent Framework, its performance is compared against individual agents, as summarized in Table 1. The LLM-based agents in this framework were implemented using the Gemini 2.5 Flash API.

\subsection{Evaluation Metrics}

\subsubsection{Levenshtein Distance}
Levenshtein distance is a widely adopted edit-distance metric used to quantify the dissimilarity between two strings. It computes the minimum number of single-character operations required to transform one string into another. The permissible operations include insertion, deletion, and substitution.

Formally, given two strings $s_1$ and $s_2$, the Levenshtein Distance $LD(s_1, s_2)$ is defined as:

\begin{equation}
LD(i,j) =
\begin{cases}
\max(i,j), & \text{if } \min(i,j) = 0 \\
\min
\begin{cases}
LD(i-1,j) + 1 \\
LD(i,j-1) + 1 \\
LD(i-1,j-1) + \delta
\end{cases}
& \text{otherwise}
\end{cases}
\end{equation}

where
\[
\delta =
\begin{cases}
0, & \text{if } s_1[i] = s_2[j] \\
1, & \text{otherwise}
\end{cases}
\]

In this study, Levenshtein Distance \cite{levenshtein1966binary} is employed to evaluate the similarity between the agent-generated fixed code and the ground-truth corrected code, thereby quantifying the syntactic correctness of the proposed repair framework.

\subsubsection{Cosine Similarity}

Cosine similarity \cite{singhal2001modern} is a vector-space similarity measure that evaluates the cosine of the angle between two non-zero vectors in a high-dimensional space. It is defined as:

\begin{equation}
\text{Cosine Similarity}(A, B) = \frac{A \cdot B}{\|A\| \, \|B\|}
\end{equation}

where $A \cdot B$ denotes the dot product of vectors $A$ and $B$, and $\|A\|$, $\|B\|$ represent their Euclidean norms.

The similarity score ranges from:
\begin{itemize}
\item $1$, indicating identical vector orientation (maximum similarity),
\item $0$, indicating orthogonality (no similarity),
\item $-1$, indicating opposite similarity.
\end{itemize}

For structural evaluation, both the agent-rectified flow graph and the ground-truth flow graph are transformed into vector representations. Cosine similarity is then computed between these vectorized representations.

A higher cosine similarity score signifies stronger structural alignment and semantic preservation between the repaired and reference flow graphs, demonstrating the effectiveness of the proposed meta-agent framework in maintaining program control-flow integrity.
\begin{table*}[htbp]
\centering
\scriptsize
\setlength{\tabcolsep}{5pt}
\renewcommand{\arraystretch}{0.9}
\caption{Comparative Analysis over the Python programs}

\resizebox{\textwidth}{!}{
\begin{tabular}{|l|c|c|c|c|c|c|}
\hline
\textbf{Name} & \multicolumn{3}{c|}{\textbf{Python LD}} & \multicolumn{3}{c|}{\textbf{Python Cosine}} \\
\cline{2-7}
 & FGDM-Standard & FGDM-COT & FGDM-TOT & FGDM-Standard & FGDM-COT & FGDM-TOT\\
\hline

Pandas-100 & 33 & 32 & 37 & 0.993 & 0.9937 & 0.8987 \\ \hline
Pandas-102 & 9 & 1 & 1 & 0.9892 & 0.9994 & 0.9994 \\ \hline
Pandas-105 & 1 & 1 & 145 & 1 & 1 & 0.9977 \\ \hline
Pandas-106 & 9 & 1 & 1 & 0.9412 & 0.9977 & 0.9977 \\ \hline
Pandas-107 & 21 & 1 & 1 & 0.9997 & 1 & 1\\ \hline
Pandas-108 & 1 & 92 & 5 & 0.9979 & 0.9784 & 0.9784 \\ \hline
Pandas-110 & 12 & 1 & 12 & 0.9753 & 0.9981 & 0.9753 \\ \hline
Pandas-111 & 21 & 6 & 21 & 0.9971 & 0.9987 & 0.9903 \\ \hline
Pandas-112 & 1 & 92 & 5 & 0.9982 & 0.8598 & 0.9812 \\ \hline
Pandas-115 & 5 & 1 & 4 & 0.984 & 0.986 & 0.9903 \\ \hline
Pandas-117 & 7 & 34 & 67 & 0.9601 & 0.8398 & 0.8251 \\ \hline
Pandas-118 & 6 & 6 & 6 & 0.9952 & 0.9952 & 0.9952 \\ \hline
Pandas-120 & 52 & 1 & 5 & 0.9932 & 0.9999 & 0.9996 \\ \hline
Pandas-121 & 13 & 1 & 21 & 0.9866 & 0.9991 & 0.9029 \\ \hline
Pandas-125 & 1 & 1 & 4 & 0.9973 & 0.9973 & 0.9818 \\ \hline

Spacy-1 & 77 & 1074 & 93 & 0.7633 & 0.4368 & 0.6742 \\ \hline
Spacy-2 & 1 & 1 & 1 & 0.9972 & 0.9972 & 0.9972 \\ \hline
Spacy-3 & 10 & 10 & 10 & 0.9865 & 0.9865 & 0.9865 \\ \hline
Spacy-4 & 60 & 17 & 1 & 0.9414 & 0.7728 & 0.9991 \\ \hline
Spacy-5 & 15 & 1 & 4 & 0.9718 & 0.9988 & 0.9921 \\ \hline
Spacy-6 & 5 & 5 & 108 & 0.9807 & 0.9807 & 0.7413 \\ \hline
Spacy-7 & 12 & 26 & 94 & 0.9942 & 0.9895 & 0.8018 \\  \hline
Spacy-8 & 1 & 1 & 1 & 0.9972 & 0.9972 & 0.9972 \\ \hline
Spacy-9 & 1 & 1 & 1 & 0.9972 & 0.9972 & 0.9972 \\ \hline
Spacy-10 & 4 & 1 & 1 & 0.9812 & 0.9973 & 0.9973 \\ \hline

Keras-10 & 1 & 1 & 153 & 1 & 1 & 0.9944 \\ \hline
Keras-12 & 1 & 1 & 1 & 0.9973 & 0.9973 & 0.9973 \\ \hline
Keras-14 & 2 & 16 & 33 & 0.9857 & 0.9619 & 0.8567 \\ \hline
Keras-15 & 1 & 1 & 48 & 0.9999 & 0.9999 & 0.9935  \\\hline
Keras-17 & 3 & 1 & 1 & 0.9878 & 0.9976 & 0.9976  \\ \hline
Keras-20 & 1 & 324 & 721 & 1 & 0.9995 & 0.8276 \\ \hline
Keras-23 & 37 & 20 & 91 & 0.6613 & 0.9951 & 0.9932 \\ \hline
Keras-27 & 1 & 1 & 15 & 0.9973 & 0.9973 & 0.9987 \\ \hline
Keras-34 & 1 & 1 & 15 & 0.9999 & 0.9999 & 0.9987 \\ \hline
Keras-40 & 1 & 1 & 55 & 0.9998 & 0.9998 & 0.976  \\ \hline

Matplotlib-10 & 4 & 1 & 15 & 0.9899 & 0.9985 & 0.9827 \\ \hline
Matplotlib-11 & 397 & 51 & 1 & 0.9406 & 0.9989 & 1 \\ \hline
Matplotlib-12 & 1 & 89 & 1 & 1 & 0.8888 & 1 \\ \hline
Matplotlib-14 & 1 & 1 & 1 & 0.998 & 0.998 & 0.998 \\ \hline
Matplotlib-17 & 1 & 1 & 36 & 0.9973 & 0.9973 & 0.9527 \\ \hline
Matplotlib-20 & 70 & 79 & 107 & 0.9116 & 0.7213 & 0.9976 \\ \hline
Matplotlib-23 & 16 & 9 & 8 & 0.9843 & 0.9617 & 0.992 \\ \hline
Matplotlib-25 & 63 & 15 & 9 & 0.9498 & 0.9658 & 0.962 \\ \hline
Matplotlib-27 & 11 & 1 & 9 & 0.9736 & 0.9978 & 0.9751 \\ \hline
Matplotlib-30 & 33 & 1 & 1 & 0.9501 & 0.9999 & 0.9999 \\ \hline

\hline
\end{tabular}
}

\end{table*}

\begin{table*}[htbp]
\centering
\scriptsize
\setlength{\tabcolsep}{5pt}
\renewcommand{\arraystretch}{0.9}
\caption{Comparative Analysis over the Python programs (contd.)}

\resizebox{\textwidth}{!}{
\begin{tabular}{|l|c|c|c|c|c|c|}
\hline
\textbf{Name} & \multicolumn{3}{c|}{\textbf{Python LD}} & \multicolumn{3}{c|}{\textbf{Python Cosine}} \\
\cline{2-7}
 & FGDM-Standard & FGDM-COT & FGDM-TOT & FGDM-Standard & FGDM-COT & FGDM-TOT\\
\hline

FastAPI-1 & 60 & 1 & 89 & 0.9994 & 1 & 0.9867 \\ \hline
FastAPI-2 & 8 & 1 & 4 & 0.9594 & 0.9982 & 0.9897 \\ \hline
FastAPI-11 & 30 & 24 & 24 & 0.8626 & 0.8381 & 0.8381 \\ \hline
FastAPI-12 & 228 & 1 & 45 & 0.4005 & 0.9999 & 0.8846 \\ \hline
FastAPI-13 & 10 & 65 & 61 & 0.9982 & 0.733 & 0.9981 \\ \hline
FastAPI-7 & 31 & 1 & 37 & 0.8719 & 0.9981 & 0.8091 \\ \hline
FastAPI-8 & 67 & 1 & 4 & 0.8138 & 0.9978 & 0.9877 \\ \hline
FastAPI-5 & 3 & 1 & 8 & 0.9934 & 0.9983 & 0.9751 \\ \hline
FastAPI-10 & 95 & 1 & 54 & 0.6403 & 0.9973 & 0.8304 \\ \hline
FastAPI-9 & 6 & 1 & 6 & 0.9818 & 0.9978 & 0.9818 \\ \hline

Ansible-1 & 4 & 1 & 1 & 0.9851 & 0.9973 & 0.9973 \\ \hline
Ansible-2 & 1 & 1 & 154 & 0.9999 & 0.9999 & 0.9532 \\ \hline
Ansible-3 & 9 & 1 & 1 & 0.9428 & 0.998 & 0.998 \\ \hline
Ansible-4 & 70 & 1 & 12 & 0.8727 & 0.9984 & 0.981 \\ \hline
Ansible-5 & 1 & 1 & 8 & 0.9997 & 0.9997 & 0.9968 \\ \hline
Ansible-6 & 131 & 1 & 1 & 0.9972 & 1 & 1 \\ \hline
Ansible-7 & 8 & 1 & 1 & 0.9867 & 0.9987 & 0.9987 \\ \hline
Ansible-8 & 1 & 1 & 4 & 0.9999 & 0.9999 & 0.9996 \\ \hline
Ansible-9 & 86 & 90 & 5 & 0.7345 & 0.9724 & 0.999 \\ \hline
 
Black-1 & 1 & 1 & 10 & 0.9993 & 0.9993 & 0.9942 \\ \hline
Black-2 & 1 & 1 & 121 & 1 & 1 & 0.9566 \\\hline
Black-3 & 12 & 1 & 12 & 0.9704 & 0.9985 & 0.9704 \\ \hline
Black-4 & 6 & 1 & 1 & 0.9703 & 0.9977 & 0.9977 \\ \hline
Black-5 & 28 & 1 & 49 & 0.9961 & 0.9999 & 0.7152 \\ \hline
Black-6 & 365 & 100 & 77 & 0.9721 & 0.9981 & 0.9892 \\ \hline
Black-7 & 60 & 1 & 1 & 0.95 & 0.9989 & 0.9989 \\ \hline
Black-8 & 1 & 32 & 17 & 0.9998 & 0.9963 & 0.9978 \\ \hline
Black-9 & 9 & 5 & 69 & 0.9593 & 0.9943 & 0.8365 \\ \hline

Luigi-1 & 6 & 6 & 6 & 0.994 & 0.994 & 0.994 \\ \hline
Luigi-2 & 1 & 1 & 77 & 0.9995 & 0.9995 & 0.7702 \\ \hline
Luigi-3 & 215 & 1 & 27 & 0.875 & 0.9993 & 0.9802 \\ \hline
Luigi-4 & 5 & 1 & 1 & 0.9703 & 0.9977 & 0.9977 \\ \hline
Luigi-5 & 1 & 1 & 55 & 1 & 1 & 0.9981 \\ \hline
Luigi-6 & 16 & 7 & 81 & 0.9997 & 0.9998 & 0.9969 \\ \hline
Luigi-7 & 13 & 1 & 1 & 0.9912 & 0.9994 & 0.9994 \\ \hline
Luigi-8 & 1 & 1 & 37 & 0.9998 & 0.9998 & 0.6076 \\ \hline
Luigi-9 & 78 & 25 & 136 & 0.9982 & 0.9989 & 0.9474 \\ \hline

Scrapy-1 & 7 & 1 & 7 & 0.9921 & 0.9993 & 0.9938 \\ \hline
Scrapy-2 & 92 & 1 & 6 & 0.9401 & 0.9993 & 0.9943 \\ \hline
Scrapy-3 & 35 & 1 & 1 & 0.9624 & 0.9988 & 0.9988 \\ \hline
Scrapy-4 & 65 & 1 & 8 & 0.9031 & 0.9987 & 0.9865 \\ \hline
Scrapy-5 & 1 & 1 & 1 & 0.9973 & 0.9973 & 0.9973 \\ \hline
Scrapy-6 & 12 & 1 & 793 & 0.9666 & 0.9973 & 0.4168 \\ \hline
Scrapy-7 & 19 & 45 & 6 & 0.9562 & 0.8994 & 0.9815 \\ \hline
Scrapy-8 & 1 & 1 & 1 & 0.9973 & 0.9973 & 0.9973 \\ \hline
Scrapy-9 & 1 & 1 & 4 & 0.9973 & 0.9973 & 0.9814 \\ \hline

Tornado-1 & 9 & 14 & 1 & 0.9846 & 0.9897 & 0.9991 \\ \hline
Tornado-2 & 56 & 1 & 26 & 0.6706 & 0.9988 & 0.714 \\ \hline
Tornado-3 & 69 & 17 & 39 & 0.9544 & 0.8966 & 0.9781 \\ \hline
Tornado-4 & 37 & 5 & 19 & 0.9958 & 0.9991 & 0.9976 \\ \hline
Tornado-5 & 17 & 1 & 16 & 0.9643 & 0.9985 & 0.9774 \\ \hline
Tornado-6 & 9 & 9 & 9 & 0.9803 & 0.9803 & 0.9803 \\ \hline
Tornado-7 & 1 & 13 & 11 & 0.9988 & 0.9508 & 0.9829 \\ \hline
Tornado-8 & 4 & 1 & 1 & 0.9815 & 0.9973 & 0.9973 \\ \hline
Tornado-9 & 1 & 1 & 4 & 0.9973 & 0.9973 & 0.9815 \\

\hline
\textbf{MEAN} & 31.28 & 24.33 & 42.2 & 0.956641 & 0.974733 & 0.951 \\
\textbf{MEDIAN} & 9 & 1 & 9 & 0.9885 & 0.9979 & 0.99115 \\
\textbf{STD DEV} & 63.7609 & 112.8355 & 109.6354 & 0.0908557 & 0.0757103 & 0.0971724 \\
\hline

\hline
\end{tabular}
}

\end{table*}

\begin{table*}[htbp]
\centering
\scriptsize
\setlength{\tabcolsep}{5pt}
\renewcommand{\arraystretch}{0.9}
\caption{Comparative Analysis over the C programs}

\resizebox{\textwidth}{!}{
\begin{tabular}{|l|c|c|c|c|c|c|}
\hline
\textbf{Name} & \multicolumn{3}{c|}{\textbf{Python LD}} & \multicolumn{3}{c|}{\textbf{Python Cosine}} \\
\cline{2-7}
 & FGDM-Standard & FGDM-COT & FGDM-TOT & FGDM-Standard & FGDM-COT & FGDM-TOT  \\
\hline

Pandas-100 & 6 & 53 & 70 & 0.996 & 0.9851 & 0.9558 \\
\hline
Pandas-102 & 6 & 16 & 6 & 0.9803 & 0.9745 & 0.9803 \\ \hline
Pandas-105 & 1 & 1 & 22 & 0.9998 & 0.9998 & 0.993 \\ \hline
Pandas-106 & 11 & 1 & 5 & 0.9835 & 0.9986 & 0.9817 \\ \hline 
Pandas-107 & 6 & 1 & 6 & 0.9964 & 0.9997 & 0.9964 \\ \hline
Pandas-108 & 23 & 1 & 95 & 0.9517 & 0.9982 & 0.9676 \\ \hline
Pandas-110 & 13 & 1 & 36 & 0.9637 & 0.9978 & 0.9506 \\ \hline
Pandas-111 & 1 & 1 & 22 & 0.9994 & 0.9994 & 0.9538 \\ \hline
Pandas-112 & 1 & 1 & 100 & 0.998 & 0.998 & 0.8378 \\ \hline
Pandas-115 & 41 & 5 & 6 & 0.9338 & 0.9741 & 0.9818 \\ \hline
Pandas-117 & 19 & 13 & 13 & 0.9765 & 0.9079 & 0.9079 \\ \hline
Pandas-118 & 7 & 1 & 7 & 0.9809 & 0.9985 & 0.9772 \\ \hline
Pandas-120 & 6 & 1 & 22 & 0.9863 & 0.9991 & 0.9378 \\ \hline
Pandas-121 & 12 & 1 & 7 & 0.9638 & 0.9985 & 0.9832 \\ \hline
Pandas-125 & 1 & 37 & 5 & 0.9966 & 0.9291 & 0.9631 \\
\hline
Spacy-1 & 24 & 8 & 14 & 0.9412 & 0.9859 & 0.9827 \\ \hline
Spacy-2 & 5 & 1 & 29 & 0.9622 & 0.9966 & 0.9186 \\  \hline
Spacy-3 & 11 & 11 & 11 & 0.9835 & 0.9835 & 0.9835 \\ \hline
Spacy-4 & 11 & 1 & 1 & 0.9772 & 0.9992 & 0.9992 \\ \hline
Spacy-5 & 1 & 1 & 4 & 0.9976 & 0.9976 & 0.9796 \\ \hline
Spacy-6 & 1 & 1 & 26 & 0.9979 & 0.9979 & 0.9606 \\ \hline
Spacy-7 & 55 & 12 & 12 & 0.9342 & 0.986 & 0.986 \\ \hline
Spacy-8 & 1 & 1 & 8 & 0.9966 & 0.9966 & 0.9694 \\ \hline
Spacy-9 & 33 & 12 & 12 & 0.9202 & 0.986 & 0.986 \\ \hline
Spacy-10 & 1 & 1 & 1 & 0.9966 & 0.9966 & 0.9966 \\ 
\hline

Keras-10 & 1 & 1 & 8 & 1 & 1 & 0.9997 \\ \hline
Keras-12 & 1 & 1 & 1 & 0.9966 & 0.9966 & 0.9966 \\ \hline
Keras-14 & 26 & 1 & 6 & 0.9514 & 0.998 & 0.9695 \\ \hline
Keras-15 & 1 & 1 & 66 & 0.9993 & 0.9993 & 0.9762 \\ \hline
Keras-17 & 1 & 1 & 30 & 0.9977 & 0.9977 & 0.9082 \\ \hline
Keras-20 & 3 & 1 & 65 & 0.9966 & 0.9996 & 0.9478 \\ \hline
Keras-23 & 6 & 1 & 19 & 0.983 & 0.9985 & 0.9494 \\ \hline
Keras-27 & 5 & 1 & 1 & 0.9622 & 0.9966 & 0.9966 \\ \hline
Keras-34 & 1 & 1 & 41 & 0.9994 & 0.9994 & 0.957 \\ \hline
Keras-40 & 11 & 1 & 11 & 0.9865 & 0.9991 & 0.9865 \\
\hline
Matplotlib-10 & 1 & 1 & 5 & 0.9978 & 0.9978 & 0.9766 \\ \hline
Matplotlib-11 & 1 & 6 & 249 & 0.9999 & 0.9987 & 0.9524 \\ \hline
Matplotlib-12 & 1 & 1 & 8 & 0.9998 & 0.9998 & 0.9978 \\ \hline
Matplotlib-14 & 5 & 1 & 1 & 0.9757 & 0.9977 & 0.9977 \\ \hline
Matplotlib-17 & 1 & 1 & 1 & 0.9966 & 0.9966 & 0.9966 \\ \hline
Matplotlib-20 & 30 & 1 & 1 & 0.9792 & 0.9993 & 0.9993 \\ \hline
Matplotlib-23 & 50 & 6 & 6 & 0.97 & 0.9841 & 0.9911 \\ \hline
Matplotlib-25 & 1 & 1 & 1 & 0.9994 & 0.9994 & 0.9994 \\ \hline
Matplotlib-27 & 11 & 1 & 11 & 0.9624 & 0.9975 & 0.9624 \\ \hline
Matplotlib-30 & 4 & 1 & 7 & 0.9924 & 0.9992 & 0.9872 \\
\hline

\end{tabular}
}
\end{table*}

\begin{table*}[htbp]
\centering
\scriptsize
\setlength{\tabcolsep}{5pt}
\renewcommand{\arraystretch}{0.9}
\caption{Comparative Analysis over the C programs (contd.)}

\resizebox{\textwidth}{!}{
\begin{tabular}{|l|c|c|c|c|c|c|}
\hline
\textbf{Name} & \multicolumn{3}{c|}{\textbf{Python LD}} & \multicolumn{3}{c|}{\textbf{Python Cosine}} \\
\cline{2-7}
 & FGDM-Standard & FGDM-COT & FGDM-TOT & FGDM-Standard & FGDM-COT & FGDM-TOT \\
\hline
Fastapi-1 & 70 & 261 & 1 & 0.9959 & 0.9813 & 0.9999 \\ \hline
Fastapi-2 & 5 & 1 & 5 & 0.984 & 0.9984 & 0.9845 \\ \hline
Fastapi-11 & 1 & 1 & 1 & 0.998 & 0.998 & 0.998 \\ \hline
Fastapi-12 & 5 & 1 & 20 & 0.9863 & 0.9988 & 0.9796 \\ \hline
Fastapi-13 & 12 & 16 & 29 & 0.9839 & 0.9897 & 0.8834 \\ \hline
Fastapi-7 & 18 & 8 & 14 & 0.9518 & 0.9776 & 0.9729 \\ \hline
Fastapi-8 & 5 & 1 & 5 & 0.9797 & 0.9979 & 0.9797 \\ \hline
Fastapi-5 & 8 & 1 & 20 & 0.9722 & 0.9982 & 0.9727 \\ \hline
Fastapi-10 & 1 & 1 & 1 & 0.9966 & 0.9966 & 0.9966 \\ \hline
Fastapi-9 & 7 & 22 & 7 & 0.9726 & 0.969 & 0.9726 \\ 
\hline

Ansible-1 & 0 & 0 & 0 & 1 & 1 & 1 \\ \hline
Ansible-2 & 9 & 0 & 5 & 0.9987 & 1 & 0.9976 \\ \hline
Ansible-3 & 0 & 0 & 10 & 1 & 1 & 0.9895 \\ \hline
Ansible-4 & 11 & 0 & 19 & 0.9941 & 1 & 0.9768 \\ \hline
Ansible-5 & 87 & 10 & 27 & 0.9715 & 0.9778 & 0.9838 \\ \hline
Ansible-6 & 0 & 0 & 18 & 1 & 1 & 0.9921 \\ \hline
Ansible-7 & 7 & 4 & 0 & 0.9894 & 0.9929 & 1 \\ \hline
Ansible-8 & 11 & 22 & 0 & 0.9993 & 0.9366 & 1 \\ \hline
Ansible-9 & 132 & 0 & 5 & 0.998 & 1 & 0.9969 \\
\hline
Black-1 & 17 & 0 & 8 & 0.9854 & 1 & 0.992 \\ \hline
Black-2 & 5 & 0 & 47 & 0.9999 & 1 & 0.9881 \\ \hline
Black-3 & 18 & 0 & 80 & 0.9747 & 1 & 0.9141 \\ \hline
Black-4 & 24 & 26 & 39 & 0.9302 & 0.953 & 0.9543 \\ \hline
Black-5 & 28 & 20 & 0 & 0.9965 & 0.9909 & 1 \\ \hline
Black-6 & 539 & 31 & 0 & 0.9825 & 0.9931 & 1 \\ \hline
Black-7 & 70 & 0 & 5 & 0.9493 & 1 & 0.9789 \\ \hline
Black-8 & 5 & 0 & 67 & 0.9993 & 1 & 0.9562 \\ \hline
Black-9 & 0 & 0 & 0 & 1 & 1 & 1 \\ 
\hline

Luigi-1 & 5 & 0 & 0 & 0.9941 & 1 & 1 \\ \hline
Luigi-2 & 619 & 2 & 85 & 0.4437 & 0.9979 & 0.9617 \\ \hline
Luigi-3 & 19 & 30 & 11 & 0.9748 & 0.9816 & 0.9902 \\ \hline
Luigi-4 & 17 & 13 & 6 & 0.9665 & 0.9814 & 0.9901 \\ \hline
Luigi-5 & 15 & 38 & 0 & 0.9996 & 0.9835 & 1 \\ \hline
Luigi-6 & 796 & 23 & 0 & 0.9808 & 0.9773 & 1 \\ \hline
Luigi-7 & 87 & 8 & 0 & 0.9796 & 0.9967 & 1 \\ \hline
Luigi-8 & 0 & 0 & 33 & 1 & 1 & 0.9751 \\ \hline
Luigi-9 & 0 & 2 & 0 & 1 & 0.9993 & 1 \\
\hline
Scrapy-1 & 6 & 20 & 0 & 0.9935 & 0.9968 & 1 \\ \hline
Scrapy-2 & 0 & 0 & 14 & 1 & 1 & 0.9822 \\ \hline
Scrapy-3 & 8 & 0 & 0 & 0.9876 & 1 & 1 \\ \hline
Scrapy-4 & 8 & 15 & 15 & 0.9857 & 0.9912 & 0.9914 \\ \hline
Scrapy-5 & 1 & 0 & 0 & 0.9982 & 1 & 1 \\ \hline
Scrapy-6 & 1 & 0 & 0 & 0.9982 & 1 & 1 \\ \hline
Scrapy-7 & 43 & 0 & 17 & 0.9363 & 1 & 0.9866 \\ \hline
Scrapy-8 & 0 & 0 & 2 & 1 & 1 & 0.9646 \\ \hline
Scrapy-9 & 0 & 0 & 0 & 1 & 1 & 1 \\
\hline
Tornado-1 & 0 & 7 & 0 & 1 & 0.9924 & 1 \\ \hline
Tornado-2 & 31 & 0 & 9 & 0.8013 & 1 & 0.9838 \\ \hline
Tornado-3 & 29 & 0 & 0 & 0.9812 & 1 & 1 \\ \hline
Tornado-4 & 4 & 0 & 28 & 0.9993 & 1 & 0.8946 \\ \hline
Tornado-5 & 23 & 0 & 0 & 0.974 & 1 & 1 \\ \hline
Tornado-6 & 17 & 40 & 0 & 0.9699 & 0.9691 & 1 \\ \hline
Tornado-7 & 10 & 0 & 7 & 0.9846 & 1 & 0.9823 \\ \hline
Tornado-8 & 0 & 0 & 22 & 1 & 1 & 0.8792 \\ \hline
Tornado-9 & 0 & 0 & 11 & 1 & 1 & 0.7646 \\
\hline

\hline
\textbf{MEAN} & 32.92 & 8.37 & 17.71 & 0.976959 & 0.992579 & 0.974748 \\
\textbf{MEDIAN} & 6 & 1 & 7 & 0.9909 & 0.99845 & 0.986 \\
\textbf{STD DEV} & 112.9312547 & 27.55294444 & 31.924562 & 0.059884917 & 0.0149821 & 0.0370745 \\
\hline

\hline
\end{tabular}
}
\end{table*}

\subsection{Comparative Analysis}

FDGM framework was evaluated using 100 real-world software programs collected from various open-source repositories, including Ansible, Black, FastAPI, Keras, Luigi, Matplotlib, Pandas, Scrapy, SpaCy, and Tornado. The benchmark includes implementations in both Python and C programming languages, thereby introducing diversity in syntactic structure and control-flow complexity.

The performance of the framework was compared against three independent prompting strategies:
\begin{enumerate}
\item Standard English Prompt
\item Chain-of-Thought (COT) pipeline
\item Tree-of-Thought (TOT) pipeline
\end{enumerate}

For comprehensive evaluation, both syntactic and semantic similarity metrics were analyzed with the following metrics as discussed in Section 5.1:
\begin{itemize}
\item Levenshtein Distance (LD)
\item Cosine Similarity
\end{itemize}

The Levenshtein distance measures the structural deviation between the generated fixes and ground-truth implementations (lower values indicate better performance), while cosine similarity measures semantic alignment (values closer to 1 indicate better similarity) (see Tables 1-4, Figure 2).

For the Python benchmark programs, the mean Levenshtein Distance (LD) values for English, COT, and TOT prompting strategies were 31.28, 24.33, and 42.20, respectively, while the median values were 9, 1, and 9. These results indicate that COT prompting achieved the lowest structural deviation, demonstrating superior syntactic performance compared to other methods.

For the C benchmark programs, the mean LD values for English, COT, and TOT prompting strategies were 32.92, 8.37, and 17.17, respectively, with median values of 6, 1, and 7. These findings further support the effectiveness of structured reasoning in improving syntactic correctness. Notably, the median LD value of 1 for COT indicates that most generated fixes closely match the ground-truth implementations.

Across both Python and C programs, COT consistently achieved the lowest median LD values, indicating superior structural alignment. English prompting showed higher structural deviations, while TOT, although beneficial for deeper reasoning, introduced variability in certain cases. The proposed flow graph-driven framework mitigates these inconsistencies by integrating multiple reasoning strategies and enforcing structural consistency.

In terms of semantic similarity, for Python programs, the mean cosine similarity values for English, COT, and TOT prompting strategies were 0.9566, 0.9747, and 0.9510, respectively, while median values were 0.9885, 0.9979, and 0.9911. These results demonstrate strong semantic alignment, with COT achieving the highest consistency.

For the C benchmark programs, the mean cosine similarity values for English, COT, and TOT prompting strategies were 0.976985, 0.992561, and 0.9974848, respectively, with median values of 0.9909, 0.9983, and 0.98525. These results indicate even stronger semantic alignment, with COT achieving near-perfect similarity scores. Although English and TOT prompts occasionally achieved comparable values, their higher structural variability suggests less stable performance.

The improved performance of the proposed framework can be attributed to its flow graph-driven integration of multiple reasoning strategies. The English prompt provides contextual understanding, COT enables step-by-step reasoning for systematic error localization and correction, and TOT explores multiple reasoning paths to enhance decision-making.

Individually, these strategies may produce incomplete or inconsistent outputs. However, the proposed framework integrates them within a unified graph-based representation, enabling validation of intermediate states, enforcement of structural consistency, and cross-verification across reasoning strategies. Additionally, the graph-based approach helps reduce hallucinations and unintended semantic deviations.

Overall, experimental results on 100 real-world Python and C programs demonstrate that the proposed flow graph-driven multi-agent framework outperforms standalone prompting strategies. It achieves lower Levenshtein Distance values, higher cosine similarity scores, improved structural consistency, and enhanced robustness against hallucinations, making it a reliable and scalable solution for automated software bug detection and repair.

% -------- CONCLUSION ---------
\section{Conclusions}

In this paper, we propose a novel multi-agent framework for software bug detection and repair based on flow graph representations and collaborative reasoning using LLMs . Our approach creates a strong synergy between the structural representation of a program and the structured reasoning capabilities of large language models. Unlike traditional methods that operate on raw source code only, the framework enables structured validation, intermediate state verification and consistency maintenance among multiple agents. The framework consists of Standard English prompting, Chain-of-Thought (COT), and Tree-of-Thought (TOT) techniques within a graph-based system architecture. The hybrid approach enables contextual understanding, systematic incremental inference and multi-branch reasoning, while mitigating the individual limitations of each prompting strategy.

We empirically evaluate the effectiveness of the proposed method on 100 real-world programs written in Python and C. Experimental results demonstrate that structured reasoning, especially the COT approach, achieves better syntactic consistency with lower average and median Levenshtein Distance values. Further, the high cosine similarity score confirms the semantic preservation. However, English prompting alone has limitations such as structural inconsistency and unstable output generation. The proposed framework tackles these challenges well by combining a graph-based structural validation and a program-wide consistency during the process of analysis and repair.

In general, the results show that adding flow graph based structural modeling to multi-step and multi-perspective reasoning can effectively improve syntactic accuracy and semantic robustness. The approach produces consistently low structural discrepancies, high contextual similarity and more resistance to hallucinations.

The study proposes a scalable and efficient multi-agent framework for intelligent and automated software repair. The suggested methodology combines program structure understanding and collaborative reasoning and offers a promising solution with high potential for practical deployment in real-world software engineering applications.

% -------- Table --------

% -------- Analysis --------
\begin{figure*}[htbp]
\centering

\vspace{-10pt} 

\begin{subfigure}{0.45\textwidth}
\includegraphics[width=\linewidth]{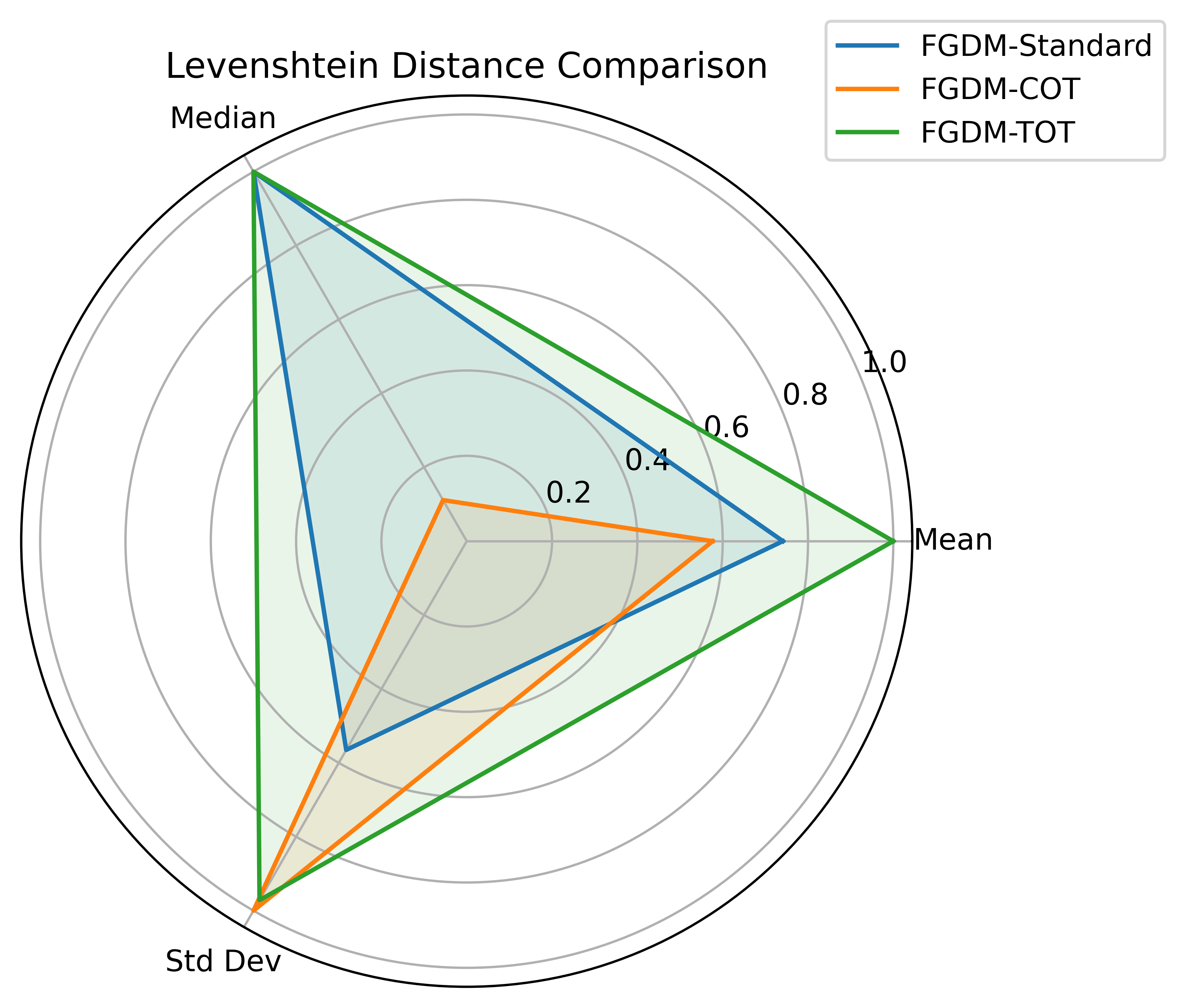}
\caption{Comparison of Levenshtein distance in Python programs}
\end{subfigure}
\hfill
\begin{subfigure}{0.45\textwidth}
\includegraphics[width=\linewidth]{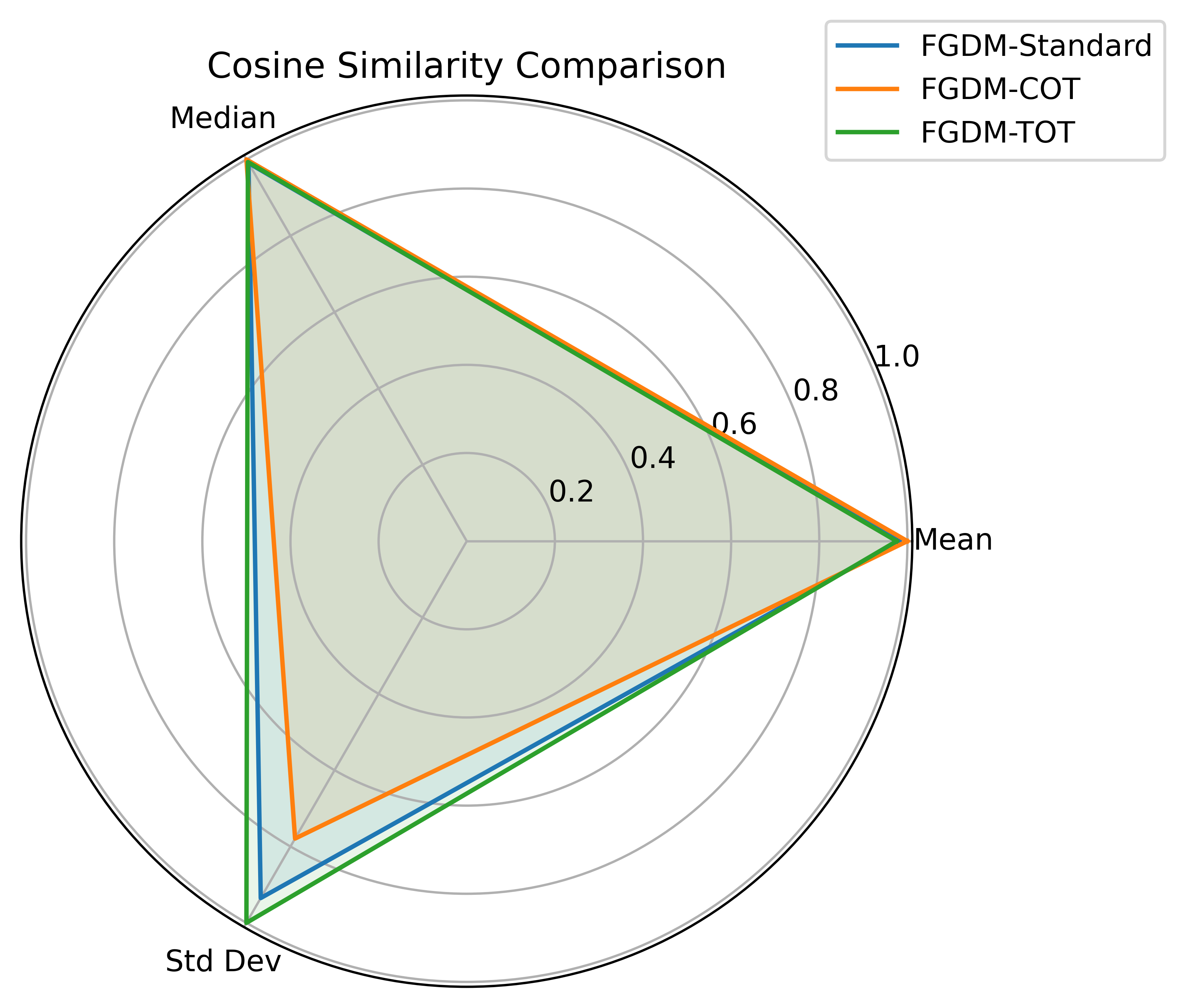}
\caption{Comparison of Cosine similarity in Python programs}
\end{subfigure}
\hfill
\begin{subfigure}{0.45\textwidth}
\includegraphics[width=\linewidth]{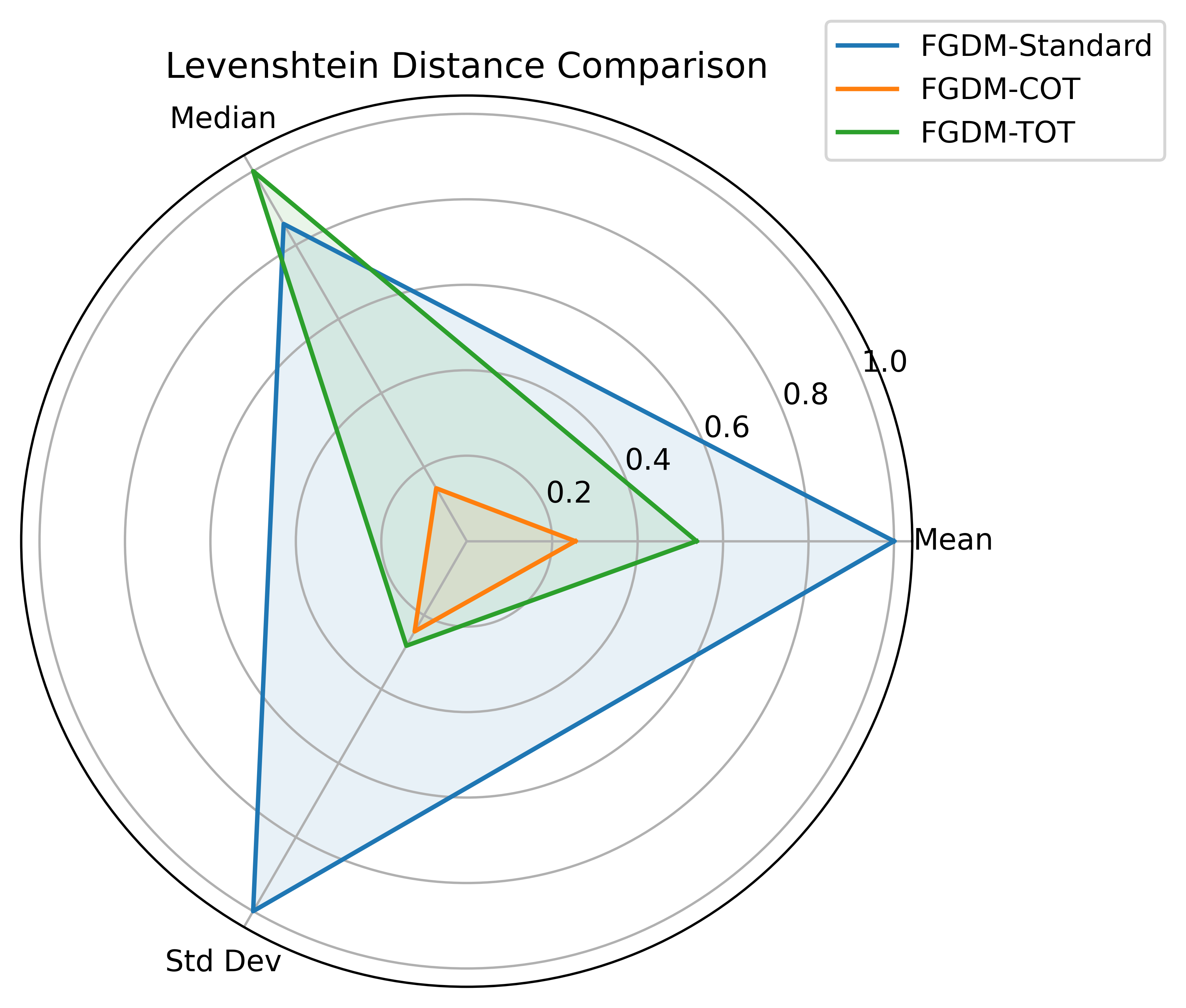}
\caption{Comparison of Levenshtein distance in C programs}
\end{subfigure}
\hfill
\begin{subfigure}{0.45\textwidth}
\includegraphics[width=\linewidth]{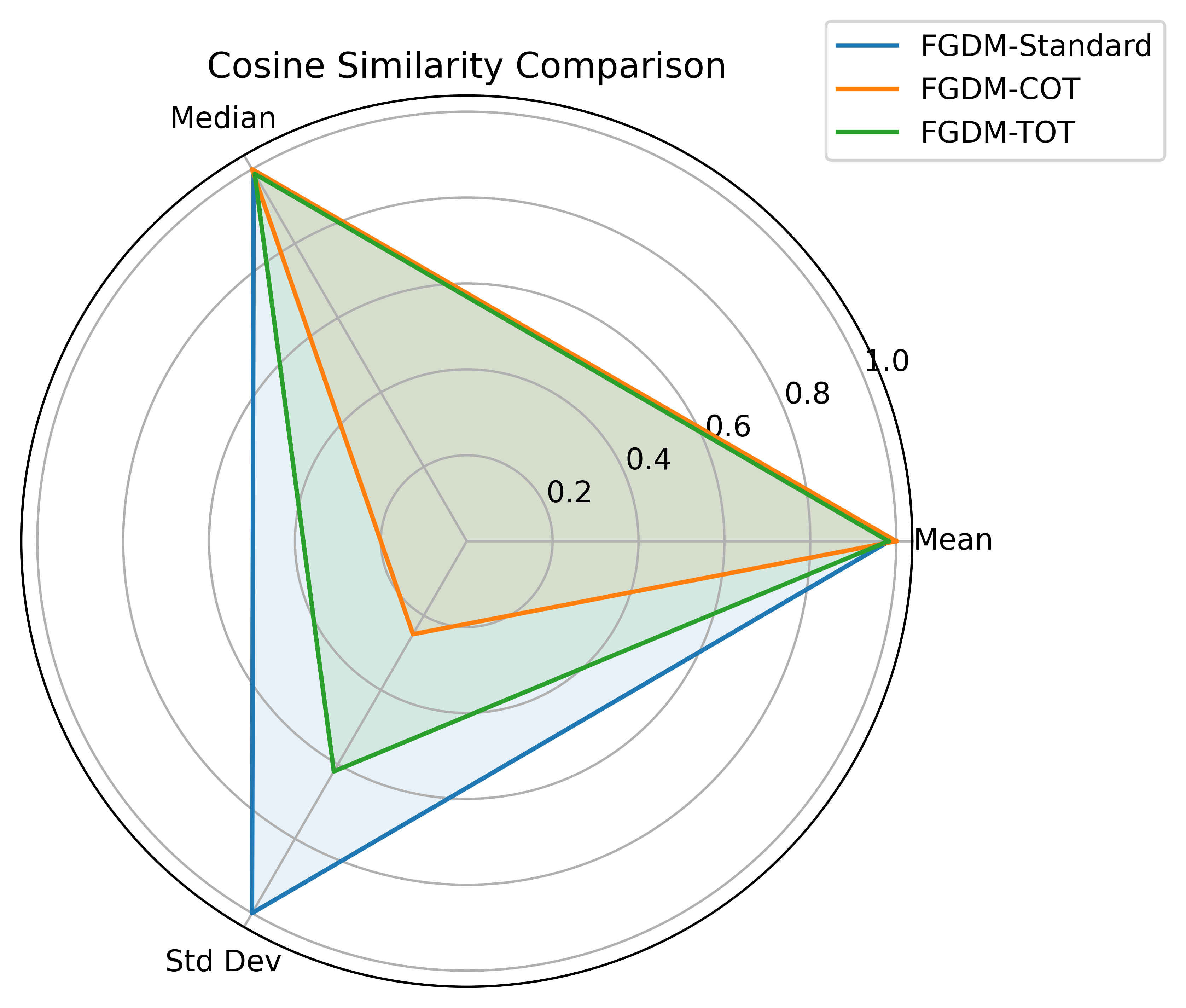}
\caption{Comparison of Cosine similarity in C programs}
\end{subfigure}

\vspace{-10pt} 

\caption{Comparison of methods}
\end{figure*}

\clearpage
% -------- REFERENCES ---------
\bibliographystyle{elsarticle-num}
\bibliography{cas-refs}

\end{document}